%% file: manuscript.tex
\documentclass[preprint,journal]{vgtc}       

\ifpdf
  \pdfoutput=1\relax                   
  \usepackage{graphicx}                
  \DeclareGraphicsExtensions{.pdf,.png,.jpg,.jpeg} 
\else
  \usepackage{graphicx}                
  \DeclareGraphicsExtensions{.eps}     
\fi%

\graphicspath{{figures/}{pictures/}{images/}{./}} 

\usepackage{enumitem}                  
\usepackage{microtype}                 
\PassOptionsToPackage{warn}{textcomp}  
\usepackage{textcomp}                  
\usepackage{mathptmx}                  
\usepackage{times}                     
\usepackage{cite}                      
\usepackage{tabu}                      
\usepackage{booktabs}                  

\onlineid{1101}

\vgtccategory{Research}
\vgtcpapertype{application/design study}

\title{IsoTrotter: Visually Guided Empirical Modelling\\ of Atmospheric Convection}

\author{Juraj P\'{a}lenik, %
		Thomas Spengler, %
		and Helwig Hauser}
\authorfooter{
\item
 Juraj P\'{a}lenik is with University of Bergen. E-mail: {juraj.palenik@uib.no}.
\item
 Thomas Spengler is with Univ.\ of Bergen. E-mail: thomas.spengler@uib.no.
\item
 Helwig Hauser is with Univ.\ of Bergen. E-mail: helwig.hauser@uib.no.
}

\shortauthortitle{P\'{a}lenik \MakeLowercase{\textit{et al.}}: IsoTrotter}

\abstract{%
Empirical models, fitted to data from observations, are often used in natural sciences to describe physical behaviour and support discoveries.
However, with more complex models, the regression of parameters quickly becomes insufficient, requiring a visual parameter space analysis to understand and optimize the models.
In this work, we present a design study for building a model describing atmospheric convection.
We present a mixed-initiative approach to visually guided modelling, integrating an interactive visual parameter space analysis with partial automatic parameter optimization. Our approach includes a new,
semi-automatic technique called \isotrotting{}, where we optimize the procedure by
navigating along isocontours of the model.
We evaluate the model with unique observational data of atmospheric convection based on flight trajectories of paragliders.
} 


\CCScatlist{ 
 \CCScat{K.6.1}{Management of Computing and Information Systems}%
{Project and People Management}{Life Cycle};
 \CCScat{K.7.m}{The Computing Profession}{Miscellaneous}{Ethics}
}

\teaser{
  \centering
\includegraphics[width=\columnwidth]{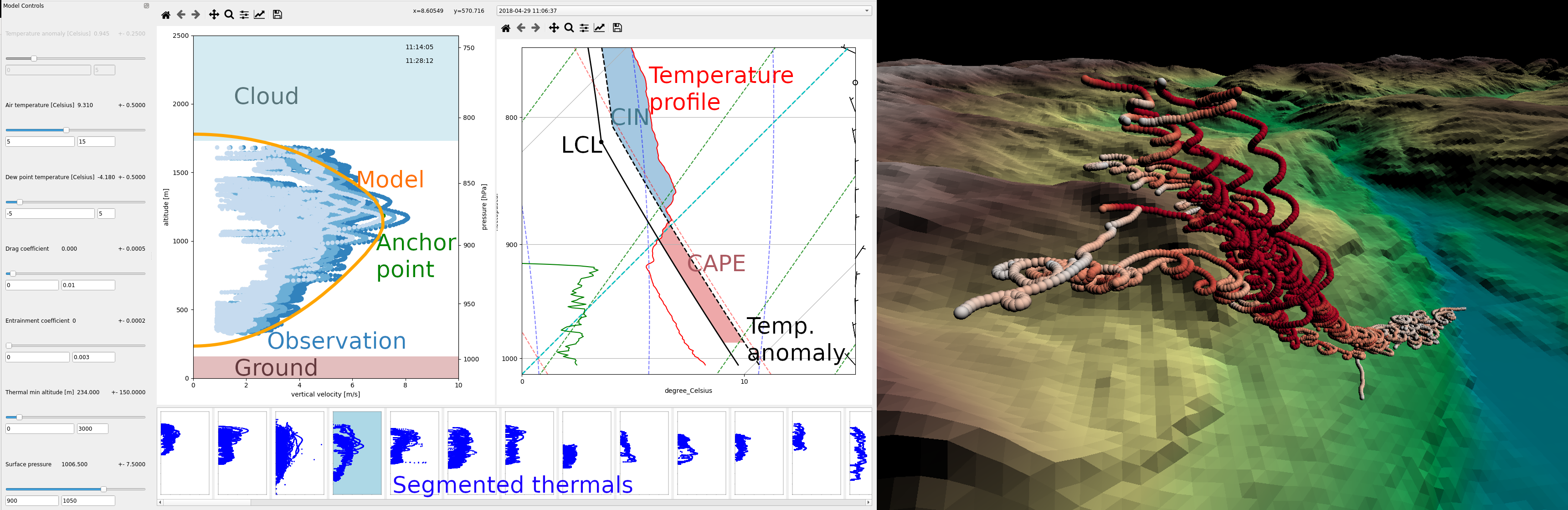}
\caption{\label{fig:teaser} \new{The convection analysis is enabled by two main views. In the left view we provide a tool for analysing the parameter-space of the model, as well as for comparing the model results to data from observations. We allow adjustment of the model parameters on the far left; the greyed-out parameter is under automatic control by the anchor point. A gallery of segmented thermals is at the bottom. On the right, a 3D view showing tracks of paragliding flights presents an instance of a thermal in its environment.}
  }%
}%

\vgtcinsertpkg

\usepackage{lipsum}
\usepackage{physics}
\usepackage{color}
\usepackage{float}
\usepackage{csquotes}
\usepackage{amsfonts}
\usepackage{epigraph}
\usepackage{subcaption}
\definecolor{darkgreen}{rgb}{0.0, 0.5, 0.16}

\newcommand{\new}[1]{{#1}}

\newcommand{\isotrotting}{IsoTrotting}

\newcommand{\R}{\mathbb{R}}

\newcommand\restr[2]{{
  \left.\kern-\nulldelimiterspace 
  #1 
  \vphantom{\big|} 
  \right|_{#2} 
  }}

\begin{document}


\firstsection{Introduction}
\maketitle

Atmospheric convection is a process in which unstable layers of air ascend and mix vertically due to their different physical properties. 
This process is responsible for cloud formation \new{as well as} transport of humidity and energy in the atmosphere. Atmospheric convection is a prominent area of research in meteorology \cite{giaiotti2007atmospheric}, where it is often explored using high-performance simulations using computational fluid dynamics~\cite{deep_convection2013} or observed data from laser light detection and ranging (LIDAR) techniques~\cite{lidar1977}.

Empirical modelling is a technique where a complex physical behaviour is described with a set of equations that are compared against experimental measurements. The equations often depend on a few free parameters that are inferred from the measurements using regression fitting. 
\new{Regression fitting can automatically select the least wrong parameters, given an error metric, once the model is established. It does not provide feedback on the suitability of a given model and the selected metric, nor does it help the scientist understand the model's behaviour.
In order to build trust and understanding of a} model with several free parameters, an extensive analysis of model behaviour under the variation of the parameters is required. This kind of analysis has previously been successfully enhanced and simplified with use of visual parameter space analysis~\cite{vPSA-framework}. 

In this work, we present a design study for empirical modelling of atmospheric convection in meteorology. In collaboration with experts in meteorology, we identify the requirements to be met for a particular combination of a model and related measurement data. Among other requirements, we recognize the importance of a parameter space analysis for successful modelling, especially the challenge when investigating the combined effects of multiple parameters. We design and implement a visual data science solution for empirical modelling and address the challenge with a novel navigation technique, focused on pairwise interactions of parameters called \isotrotting{}. This technique enables rapid identification of complex behaviour of empirical models. 

For validation of the model we use an innovative data source. The local realizations of atmospheric convection are extracted from a freely available database of tracks of paragliding flight trajectories. 
Once airborne, these engine-less air-crafts are only able to gain altitude using regions of \new{buoyant, ascending} air in the atmosphere (thermals), \new{representing} the upward moving branch of \new{convective overturning in the atmosphere}. 
\new{Quantifying the} lift experienced by these aircraft, combined with its flight properties, we \new{are} able to infer the local strength of the convection. 

Our solution helps the user in three stages: \vspace{-1em}
\begin{enumerate}[itemsep=-0.5ex]
\item Processing the aircraft trajectories into measured quantities accounting for different modes of flight
\item Using techniques of visual parameter space exploration to map model dependencies 
\item Summarizing the findings by communicating the uncertainties of the model and observations
\end{enumerate}

\section{Related Work}

\new{Hsieh et al.~\cite{Hsieh2011} addressed a post flight analysis of tracks of gliders. They propose a set of visualizations for analysing derived quantities, including thermal lifts. Our work shares parts of the data processing approach, though in addition we device an empirical model of convection with our focus being on the visual analysis of the parameter space of the model. The processing of thermals in our work is also distinct by aggregating the data from co-located flights.}

Several tools and techniques for data analysis have been developed for meteorology~\cite{MetVisSurvey2018} and visualization for physical sciences~\cite{PhysVisSurvey2012}.
However, visually building empirical models in meteorology has so far not been addressed.
The speciality of our application is that it allows the use of visualization in an interactive visual analysis framework for meteorological research. 

There are a number of notable works that are concerned with the validation of complex models and simulations. For example, Ahrens et al. focus on verifying code in scientific simulations~\cite{VerifyingSim}.
Dransch et al.~\cite{AssessingSims2010} focus on confidence in complex scientific simulations, and Unger et al.~\cite{Dransch2012} present a concept for validating geoscientific simulations.
The book \enquote{Assessing the Reliability of Complex Models}~\cite{national2012assessing} goes in a similar direction, and Kehrer et al.~\cite{Kehrer2008} focus on hypothesis generation in climate data.
While these works are valuable resources, they have not addressed how one can combine empirical models and real-world observations, which poses its own set of challenges.

From a methodological point of view, we follow the visual parameter space analysis framework by Sedlmair et al.~\cite{vPSA-framework}. The authors analyse 21 prominent publications in visual parameter space analysis. Our approach shares a navigation strategy with the works of Unger et al.~\cite{Dransch2012} and Brecheisen et al.~\cite{dti2009}. The work by Guo et al. on multivariate linear fitting~\cite{linear2009} implements the idea of empirical modelling, though only linear models are considered.
Other works concerned with regression and fitting are the work of M\"uhlbacher and Piringer on partition based model building~\cite{Piringer2013partition}. While their work is on the opposite end of the spectrum of empirical modelling, constructing the model from data alone, we first derive a model based on the underlying physics with variable parameters that are not sufficiently constrained by the theory.

The idea of exploring a high-dimensional space through anchor points can be attributed to van Wijk et al. in Hyperslice~\cite{Hyperslice}, later revisited by Torsney-Weir et al. in Sliceplorer~\cite{Sliceplorer}. In their respective works the authors compute axis-aligned slices of a high dimensional scalar-valued function. We, on the other hand, use the anchor point to define an iso-contour in the parameter space.
A good overview of isosurfaces and their applications can be found in the book \enquote{Isosurfaces} by Rephael Wenger~\cite{wenger2013isosurfaces}.
The iso-contour in our case, however, is not an object of rendering as in volume visualization. Our iso-contour is a conceptual area in the parameter space, which provides approximately the same results of the simulation runs under specific criteria. 

A common denominator when exploring a parameter space is dealing with uncertainty, which is summarised in the book \enquote{Scientific Visualization}~\cite{hansen2014scientific}. Exploring the uncertainty in model simulations is often combined with a sensitivity analysis~\cite{pointwise_sensitivity2011}, sometimes analysing sensitivities in ensemble simulations~\cite{ensembleSensitivity2019}.
We share the aim of explaining a computational model with M\"uhlbacher et al.~\cite{BlackBox2014}, though they are focusing on integrating existing libraries, whereas we devise and examine a custom model.
Our \isotrotting{} approach to parameter space exploration can be compared to work of Lindow et al.~\cite{hege2012parameters}, where the authors re-parametrize the input to achieve perceptually linear outcomes. \new{Our work is concerned with a different task of navigating regions of parameter spaces that result in similar outcomes.}

\new{
The use of empirical models in computer science is called \emph{surrogate modelling} and has been recently dominated by machine learning approaches. For example, Couckuyt et al.~\cite{sumo2013} presented SUMO toolbox, an open-source implementation of machine learning algorithms for automatic building of surrogate models. 
No automatic model building is possible in our case, since we build a surrogate model for a natural phenomenon.
Lampe et al.~\cite{lampe11modelbuilding} propose a visualization approach for interactive model prototyping based on residual analysis. Their work is concerned with statistical modelling, where ours is concerned with dynamic systems.}

\new{
Queipo et al.~\cite{QUEIPO2005} review the surrogate-based analysis and optimization in aerospace industry, 
Jin et al.~\cite{metamodelling2001} provide a comparative study of four metamodelling techniques and
Eldred et al.~\cite{Eldred2006} describe algorithmic techniques relevant for surrogate optimization. 
The optimization in this context refers to the optimization of the original simulation, whereas we optimize the surrogate (empirical) model.}

\new{
Surrogate models are also used for steering of scientific simulations.
Butnaru et al.~\cite{Butnaru2012} utilize parallel implementation of surrogate modelling for interactive steering of large scientific simulations. 
Matkovi\'{c} et al.~\cite{Matkovic2011, Matkovic2014} describe visual interactive steering of complex simulations using analytical models in the simulation space.
The steering of simulations is not applicable to our problem as opposed to the interaction wiht the surrogate models. We propose a novel interaction technique.}

\new{
While the goal of surrogate models is to speed up the computation of large simulations to enable optimization as well as interactive analysis and steering of simulations, our work contributes to the interaction with the parameter space of the empirical/surrogate model itself and can be viewed as complementary to traditional surrogate modelling.}

%
\section{On Visually--Aided Empirical Modelling}
%
We have adopted a task-centered approach in our collaboration with the meteorologists, involving a rapid prototyping process with frequent meetings and discussions to define the tasks to be addressed by the visualization as described by Tamara Munzner~\cite{munzner2014visualization}.
In this section, we describe the process and provide the results of our task analysis.

\subsection{Objectives}
The objectives of our collaboration were twofold. First, the dataset had to be explored and analysed. Second, the model had to be designed, optimized, and evaluated against the observed data.

1.) As the paragliding dataset is new to the domain experts, the first objective was to gain insight into the information contained in the data: to determine the implications of using aircraft trajectories for sampling of the atmospheric convection; to understand the sources of uncertainties and the accuracy and granularity of the data; and to propose necessary filtering and post-processing steps. 

2.) The second objective was to devise and interpret the model and become aware of its limitations. While the kernel of the model is based on principal physics of the atmosphere, empirical elements remain, as some processes need to be approximated due to their complex behaviour. Understanding the physicality and limitations of the model is essential for the model validation process.
While the physical interpretation of the model is straightforward, the biggest challenge in the model analysis is associated with the large and complex parameter space. Not only has the model several parameters, but the interplay of these parameters introduces another level of complexity.

\subsection{Requirements}

From our analysis of the problem, we were able to identify five analytical requirements (AR) and two visualization requirements (VR). 
\begin{description}
\item[AR0 -- Establishing the rate of sink for the aircraft.\hfill]

Aircrafts without an engine are constantly sinking, which means that the observed vertical velocity is not the actual vertical velocity of the air. In order to retrieve the actual vertical velocity of the air from the aircraft trajectory, one needs to account for the rate of sink.

\item[AR1 -- Segmenting the convective mode of the trajectories.]

Pilots navigate the air space along a desired track looking for the areas of rising air. There are many manoeuvres that might be executed, but from an analytical point of view, we can distinguish three different modes of flight: 1.) Free flight, when the pilot is flying straight towards a destination; 2.) Thermal flight, when the pilots are \new{circling in an area of rising air with the aim to gain altitude}; 3.) Soaring, when the pilot uses lift associated with the deflection of wind against a mountain ridge. For the purposes of analyzing atmospheric convection, only the second mode is of interest.

\item[AR2 -- Spatio temporal clustering of thermal segments.]

Depending on the size of the convection and the quality of segmentation, several segments coming from the same or different trajectories might be co-located and contribute to the characterisation of a single thermal. These segments need to be identified and clustered to provide information about a single realisation of the convection model.

\item[VR0 -- Visual inspection of the segmented thermals.]

The information about a single thermal can be manually inspected to gain confidence about the automatic processing and account for possible errors.

\item[AR3 -- Analysis of model parameters.]

The model uses a computational method to turn input parameters into output values. Understanding the effect of the input parameters on the output is the most important part of the model building. \new{This goes far beyond a simple fitting of parameters to the observed data. The domain experts are interested in several aspects including: isolated effects of each parameter; combined effects of multiple parameters; invariants in the parameter space, i.e., exploring alternative parameter settings for identical/similar outcomes. An example of practical questions would be: \emph{How would the results differ had the day been warmer? What parameters yield the same maximum ascent? What combination of parameters yields maximum ascent at a given height? Are there redundant parameters?}}

Among the identified requirements, \textbf{AR3} stands out as the most challenging. Automatic regression fitting cannot answer the above questions, which is why we involve the expert and support them with an interactive visual parameter space analysis. 

\item[VR1 -- Visualization of input parameters of the model.]

Translating the model parameters into a visual representation provides the domain expert with more intuition for the functioning of the model and the expected behaviour. The ability of the domain expert to make predictions about the model outcome is essential for the model validation and optimization.

\item[AR4 -- The model evaluation.\\]

One needs to be able to asses the quality and uncertainty of both the model as well as the observations and the sensitivity of the model to the input parameters.
\end{description}

\subsection{Visual Parameter Space Analysis}\label{sec:vpsa}
We find that the best way to explain the analysis of the model parameters (\textbf{AR3}) in this work is by using the vocabulary and taxonomy defined in the work of Sedlmair et al.~\cite{vPSA-framework} on visual parameter space analysis (vPSA framework from now onward). 
The equations of the model can be found in the supplementary material, here we provide a gist of the model behaviour and outline our design decisions using the language of the vPSA framework.

\subsubsection{I/O Parameters}
\label{sec:params}
The vPSA framework distinguishes between two \emph{types} of inputs and outputs:  \emph{multi-variate/multi-dimensional} objects and \emph{complex objects}. A multi-variate input is for example a simple variable, whereas a complex object can be an image.

The framework further defines three \emph{classes} of input parameters: 1.)~\emph{control parameters} -- the analysed parameters; 2.) \emph{environmental parameters} -- parameters measured outside of the model and hence treated as random variables; 3.) \emph{model parameters} -- parameters controlling the inner workings of the model.

The model in our case is a system of partial differential equations describing the vertical motion of air in terms of buoyancy induced by a temperature perturbation. 
The result of integrating the model yields a vertical velocity profile $w(z)$, where $w$ is the vertical velocity at a given altitude ($z$). The profile represents the strength of the updraft and is the meeting point for the observed data.
In the language of the theoretical framework~\cite{vPSA-framework}, the output would be considered a complex object. In the mathematical language this complex object is a one dimensional scalar valued function $w(z) : \R \to \R$, where vertical velocity is assigned to each altitude.

\new{The reader might wish to think about the treatment of the thermals in our model as a hot air balloon simulation. The balloon is filled with air slightly warmer than the environment, which at the same pressure has lower density than the surrounding air, resulting in positive lift. The upward motion of the balloon is hindered by drag. As the thermal is not enclosed in any fabric, the balloon can be considered to be perforated, allowing for air inside and outside of the ballon to be exchanged, which we refer to in the model as \emph{entrainment}. The result of the simulation is the balloon's vertical velocity. Any sideways motion due to wind is disregarded and mostly irrelevant to the problem.}

There are several input parameters for the model.
There is a single complex environmental parameter called the air profile, which is obtained from a radiosonde measurement from a nearby station. A radiosonde is a commonly used device attached to a helium-filled balloon taking meteorological observations while ascending through the atmosphere.
Mathematically, this input parameter is a multi-dimensional function $\phi(z): \R \to \R^n$, which provides a series of observations for each sample at different altitudes. These observations include: pressure, temperature, dew point temperature, wind speed, and wind direction.
The rest of the parameters are scalar values, i.e., multivariate parameters in the language of the vPSA framework.
Their categorization would at first sight be either \emph{environmental parameters} or \emph{model parameters}. However, as all these parameters are of primary interest to the domain experts and are thus subject to the examination, their treatment corresponds to that of \emph{control parameters}.
 
\begin{description}
\item[Air profile.] A complex environmental parameter describing the observation at the nearest meteorological station. \new{In our case, the observation is obtained using a radiosonde measurement released at 12-hour intervals from Sola airport near Stavanger (200 km away from the paragliding area). Due to the temporal and spatial difference between the air profile measurement and the paragliding flights, a part of the air temperature profile in the lower altitudes is replaced by an explicit dependency based on dry adiabatic processes.}
\begin{equation}\label{eq:temp}
    \theta(P) = T\left(\frac{P_0}{P}\right)^{\frac{R}{C_p}}
\end{equation}
\new{where $\theta$ is the potential temperature as a function of pressure $P$, $T$ is the two-metre temperature, $P_0$ is the reference pressure and the exponent is a constant $\frac{R}{C_p} = 0.286$. The equation for potential temperature gives the temperature a parcel of dry air would have if adiabatically moved to the altitude of the reference pressure $P_0$, i.e., without any energy exchange with the surroundings. While ascending, the temperature of the air parcel decreases as pressure is decreasing.}

\item[Surface pressure.] A control parameter responsible for matching the difference between the altitude of the observation of the air profile and the location of the observed and simulated convection. For simplicity, we set the reference pressure $P_0$ to be equal to the surface pressure.

\item[Two-metre temperature.] A control parameter which can be estimated by an on-site measurement, corresponding to the air temperature at the location of the thermal simulation/observation.

\item[Two-metre dew point temperature.] A control parameter which can be estimated by an on-site measurement, indicating the amount of humidity at the location of the thermal. \new{The dew point temperature is a computed value at which the air vapour would start to condense given a cooling under isobaric (constant pressure) conditions. It is needed for the computation of the altitude where clouds start to develop, called the \emph{lifted condensation level (LCL)} or the \emph{cloud base}.}

\item[Temperature anomaly.] The main parameter controlling the perturbation of the observed state. The temperature anomaly captures the temperature difference between the thermal and the surrounding air. It is theoretically possible to observe the temperature anomaly, though this data is difficult to obtain without advanced observational strategies that are not readily available.

\item[Drag coefficient.] A model parameter controlling the intensity of the force acting against the updraft. \new{It represents the environment's friction forces slowing down the ascent of the rising air.} This parameter is considered a control parameter associated with the model building and corresponds to the parameter $\alpha$ in the equations in the supplementary material. 

\item[Entrainment coefficient.] A model parameter controlling the dissipation of the temperature perturbation. \new{If we imagine the thermal as a hot air balloon, this parameter controls how perforated the balloon is.} This parameter is considered a control parameter associated with the model building, corresponding to the parameter $\gamma$ in the equations in the supplementary material. 

\item[Thermal minimal altitude.] A control parameter corresponding to the spatial configuration of the thermal, determining the altitude at which the thermal first develops. It is connected to the topographic data of the location of the thermal.

\end{description}

\subsubsection{Sampling}
The evaluation of a single model run requires 0.1 second, which means that we do not need to pre-sample the model and that we can perform the application interactively, even including a full model evaluation. We have performed a cluster analysis on the results of sampled model runs using latin hypercube sampling~\cite{latin_hypercube1987} (see below). Should the running costs of the model evaluation increase with future improvements of the model, an appropriate sampling technique will be utilized according to recommendations from the vPSA framework~\cite{vPSA-framework}.

\subsubsection{Navigation strategies}
\label{sec:strategy}
The three navigation strategies described by the vPSA framework~\cite{vPSA-framework} are 1.) Informed Trial and Error; 2.) Local-to-Global; and 3.) Global-to-Local.
It could be argued that any application that provides users with a possibility to change the input parameters and to display the result automatically also supports the first navigation strategy. We were therefore considering only which of the other two would be more beneficial in our case. In our design study, we have identified the user's need for parameter and sensitivity analysis (\textbf{AR3}, \textbf{AR4}). We have also discovered that the model outputs vary smoothly in the parameter space, and that cluster analysis on sampled model runs did not yield distinct modes as in the work on explosions by Brucker and M\"oller~\cite{Bruckner2010}.

On the other hand, there are a couple of reasons for using the Local-to-Global navigation strategy.
First, the empirical nature of this work with the possibility of several parameters being measured, or at least partially estimated, provides the user with good initial parameter settings. 
This restriction of the parameter space comes from physical realism. The user is not interested in the nuances of unphysical behaviour of the model in unphysical situations. The user, rather, is interested in the qualities of the model when the model parameters are close to the observed values, as well as the change of the behaviour associated with slight variations.

Further reasons for using a Local-to-Global navigation strategy are the parameter analysis and model evaluation requirements of \textbf{AR3} and \textbf{AR4}. The user spends much more time \new{exploring alternate parameter configurations to evaluate the model's sensitivity} rather than looking for a global behaviour approximating the data, especially if they are provided with on-site measured estimates of parameter values. 
\new{Most difficult to judge are the combined effects of multiple parameters on the model. A change in a single parameter, the drag for example, will have a dampening effect on the whole profile and the vertical velocity will be smaller. This, however, is not relevant for the domain expert who is interested in how the other parameters need to change, if the drag is bigger, such that the profile still corresponds to the observed profile.}

For these reasons we tried to find a suitable Local-to-Global navigation strategy. However, out of 21 analysed papers in the vPSA framework, only three employed this strategy. Each of these works uses a custom-tailored navigation strategy, neither of which was suitable for our model. As we have not found any suitable technique in the literature, that would satisfy the domain experts special interest in the mutual interaction of parameters, we developed a novel navigation technique called \isotrotting{}.

\section{IsoTrotting}
\label{sec:iso-trotting}

\isotrotting{} is an interaction technique that enables the navigation of the parameter space of scalar models in a new way. The technique is most useful when the parameter space is being explored for alternative solutions and it solves the problem of analysing the interplay between two input parameters. 
Without a proper navigation technique, analysing the joint effect of two parameters is a tedious task, where the user has to manually browse through parameter values which are not relevant and gets distracted from the analysis task.
In order to address this manual and repetitive readjustment of parameters, we propose a novel technique to constrain the parameter space using anchor points in the output domain.

\subsection{Formal Definition}
Following the black-box metaphor of the model from the vPSA framework\cite{vPSA-framework}, we describe \isotrotting{} as a general technique to explore iso-structures in parameter spaces.

The black-box is assumed to be a scalar model $F$ with $n$ parameters and $m$ variables
\begin{equation}\label{eq:one}
\begin{array}{rl}
F:   & \R^{n+m} \to  \R  \\
     & (u_1, \dots, u_n, x_1 \dots, x_m)  \mapsto \quad  y
\end{array}
\end{equation}

\begin{figure}[bt]
\includegraphics[width=0.49\columnwidth]{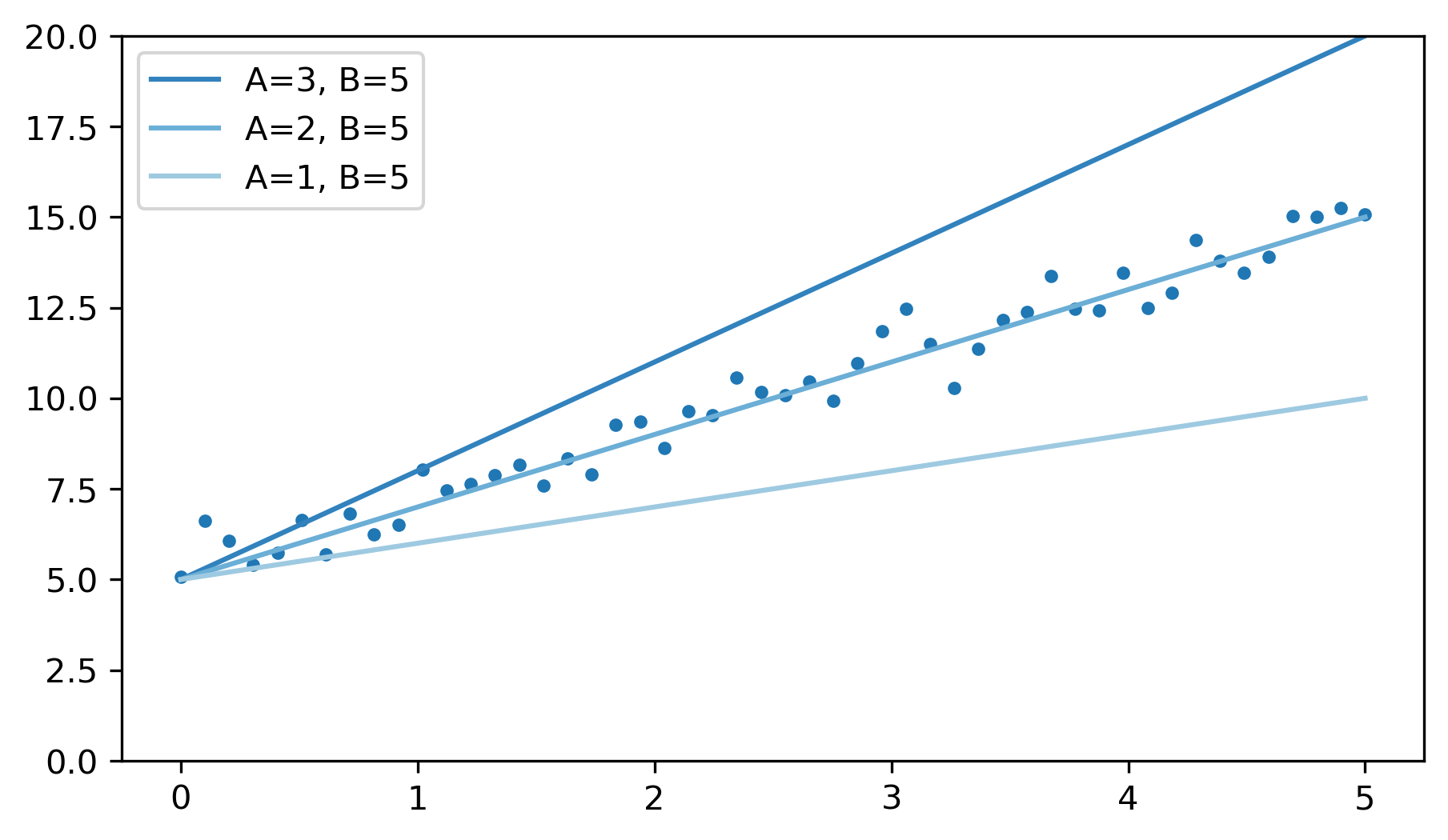}
\includegraphics[width=0.49\columnwidth]{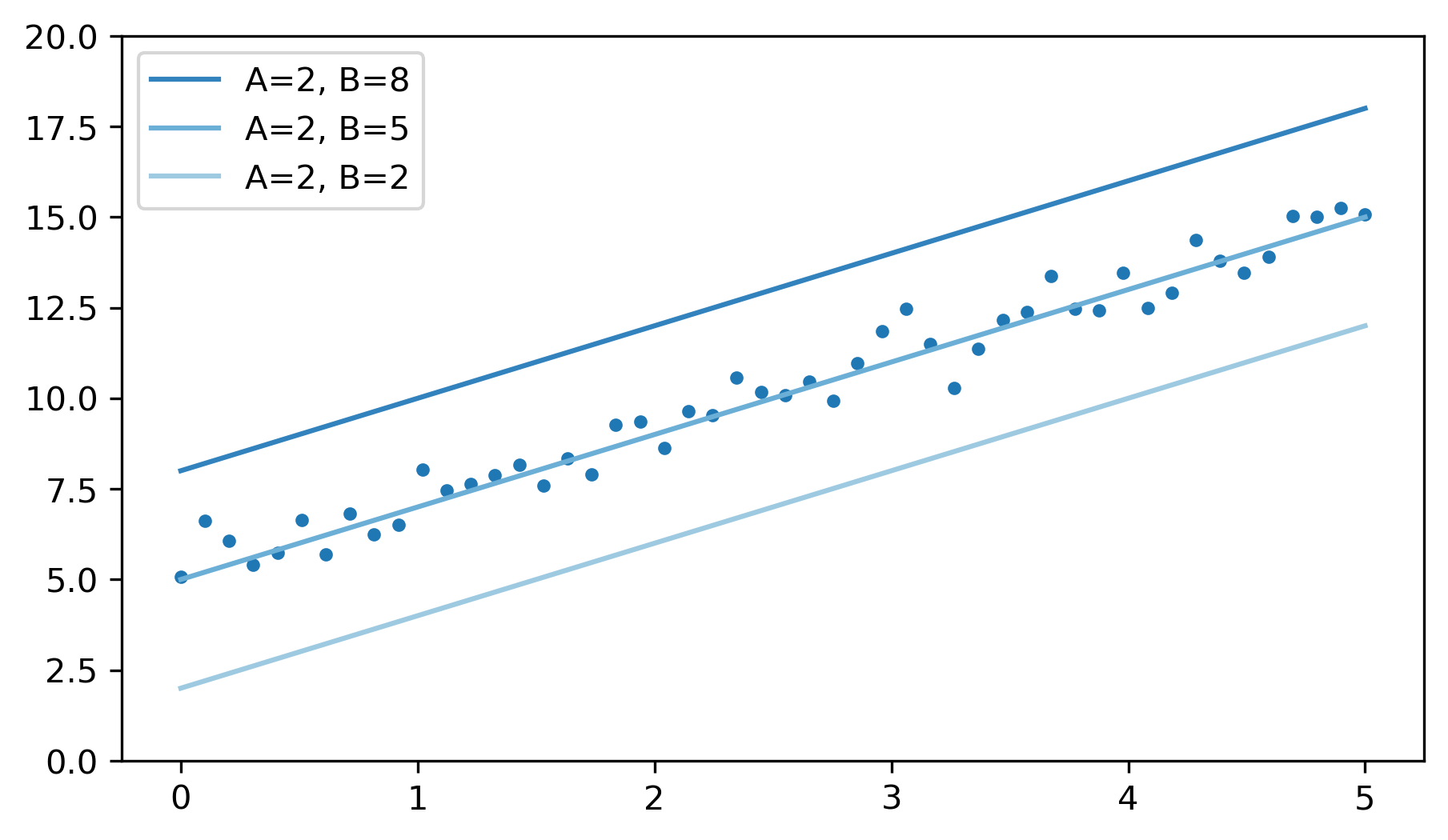}
\includegraphics[width=0.49\columnwidth]{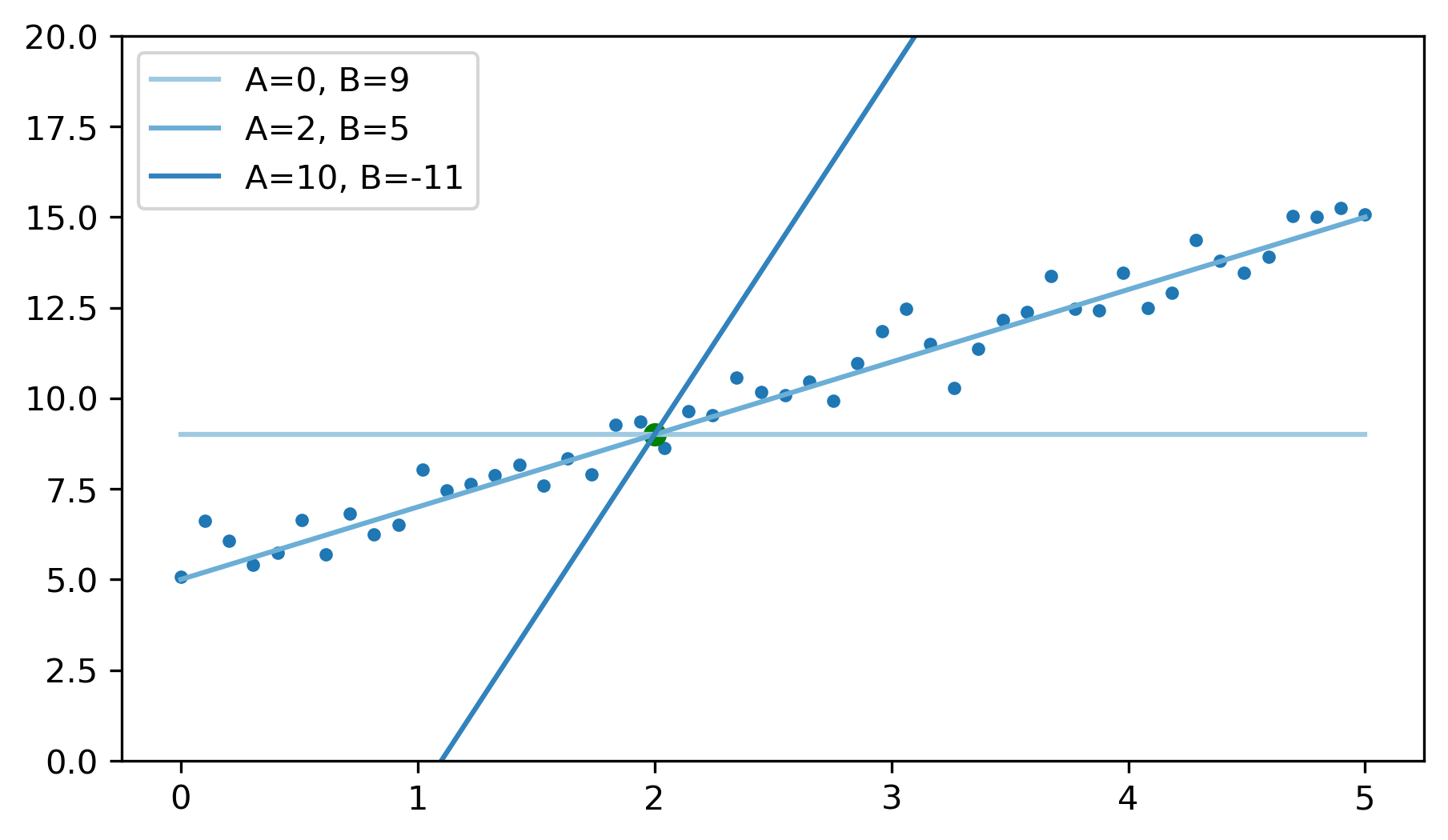}
\includegraphics[width=0.49\columnwidth]{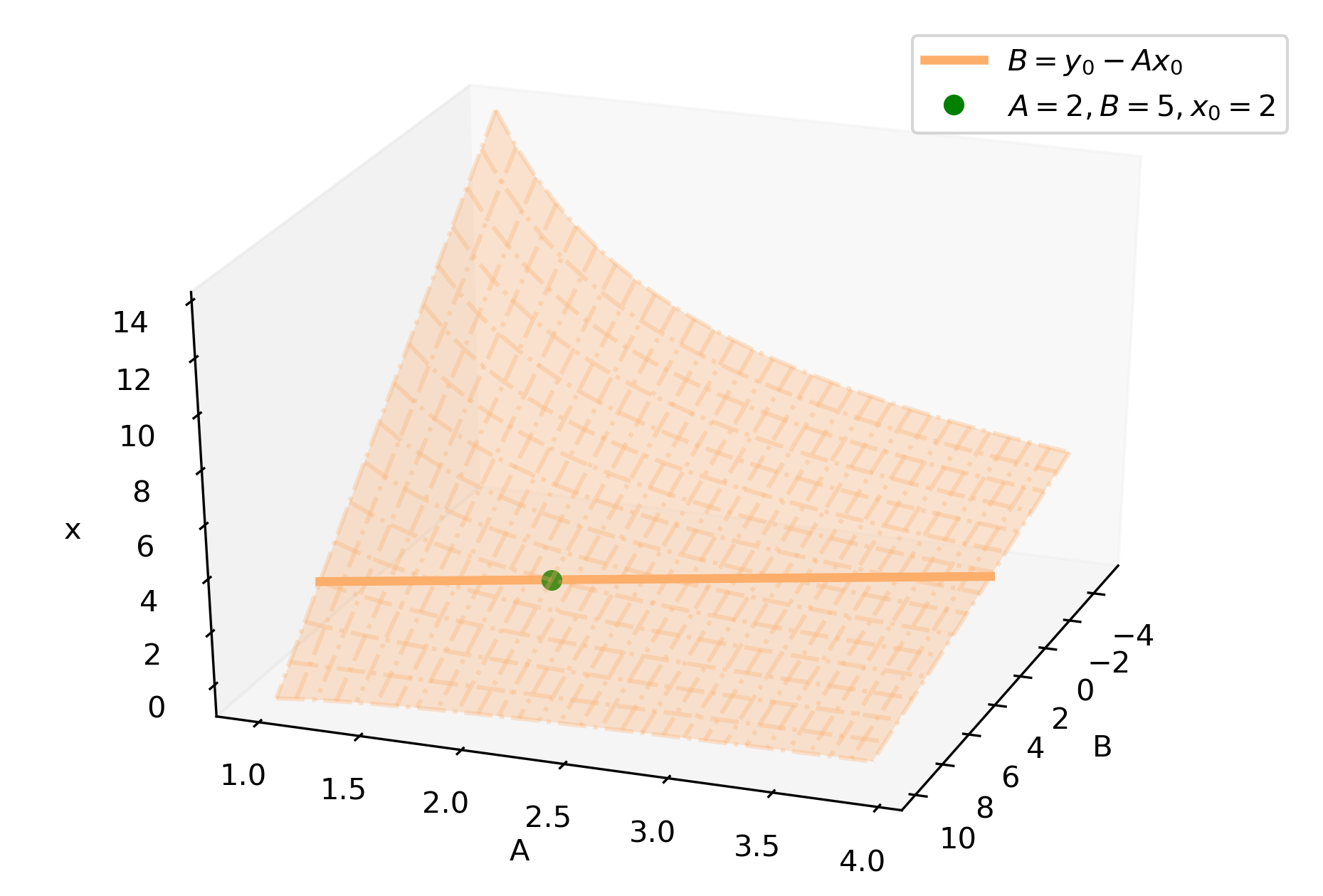}

\caption{\label{fig:synthetic} A synthetic example of the \isotrotting{} principle. A linear model $y=Ax+B$ is analysed, with two parameters $A$ and $B$. Top-Left: The variation of parameter $A$ is responsible for change in the slope of the model. Top-Right: Variation of the parameter $B$ is responsible for offset of the model. Bottom-Left: Placing an anchor point in the range of the model defines an isosurface in the parameter space. Bottom-Right: The isosurface in the (A,B,x)-space intersected with $x=x_0$ plane gives the orange isoline. \new{\isotrotting{} is achieved by traversing along the isoline in the (A,B)-space, which restricts the parameter exploration to values such that the model will pass through the anchor point in the (x,y)-space.}  }
\end{figure}

A particular example would be a linear model: $y = Ax + B$. The model parameters are $A$ and $B$, with $x$ being the input variable and $y(A,B,x)$ being the output variable. The realization of the model, given parameters $A$ and $B$ would be a linear function $g(x) = Ax + B$ (see \autoref{fig:synthetic}). The informed trial and error strategy would be achieved by adjusting each of the input parameters independently, traversing the parameter space along the orthogonal parameter axes. In this example changing the parameter $A$ would result in changing the slope of the function $g$ (\autoref{fig:synthetic}, top-left) and interactive adjustment of the parameter $B$ would result in a movement of the function $g$ along the $y$-axis (\autoref{fig:synthetic}, top-right). 

\new{
With \isotrotting{}, instead of traversing the parameter space along the parameter axes, we traverse the parameter space along the model's iso-contours. First an iso-value $y=y_0$ is selected, which defines a $(n+m-1)$-dimensional hyper-isosurface. This hyper-isosurface is intersected with $n+m-1$ hyper-planes, specifying the variables $x_i=x_{i,0}$ and the parameters $u_j = u_{j,0}$, respectively. Each hyper-plane intersection will reduce the dimensionality of the hyper-surface by one. By specifying $n+m-1$ hyper-planes a point in the parameter space lying on the hyper-isosurface is defined. This leaves one parameter/variable, whose value is bound on the iso-surface. Varying any of the positions of the hyper-planes makes this point in the parameter space trace a curve lying on the hyper-isosurface.}

\new{
Varying the isovalue $y_0$ is also possible and results in a one dimensional fitting of the bound parameter to the specified output. This will not keep the point on the same isosurface, but instead create a new isosurface which, in general, can be topologically different from the previous one, or it might not exist at all.}

\new{
In theory, the intersection of hyper-planes with the isosurface can be empty, unique, or non-unique (multivalued). Avoiding empty intersections is easily achieved by starting from a chosen point in the parameter space and its value as the isovalue. Non-unique solutions are a result of topological properties of the implicit surfaces, study of which is out of scope of this paper. Instead, we propose exploring the isocontours locally, using small continuous variations of the parameters, which will mitigate the complications of the non-unique global solutions. This does not take away from the usefulness of our method, since any connected smooth manifold can be traversed by local variations.}

\new{
In the example of the linear model, we would fix a value $y=y_0$, which defines a 2D isosurface in the three-dimensional $(A,B,x)$-space (\autoref{fig:synthetic}, bottom-right). Specifying a hyper-plane $x=x_0$ defines the intersection curve in the $(A,B)$-space by the implicit equation $F(A,B) = A x_0 + B = y_0$. This can be solved explicitly for one of the parameters as $B(A) = y_0 - Ax_0$. Changing the parameter $A$ then dictates the change in the parameter $B$, tracing a line in the ($A,B$)-space, rotating the line about the anchor point in the ($x,y$)-space (\autoref{fig:synthetic}, bottom-left).
}

The above example is obviously simplified for explanation purposes, to illustrate the principle of the method. The benefit of \isotrotting{} is more profound when the model gets more complicated, for example, when a numerical solution of differential equation with several parameters is explored.

\subsection{Implementation}
\new{
In practice, the method works as follows: Our one-dimensional model yields a diagram in the ($x,y$)-space, corresponding to the currently-set parameters. Each of the parameters specifies a hyper-plane in the parameter space.
Instead of presenting hyper-planes to the user, we provide the notion of an anchor point. By placing an anchor point $p = (x_0, y_0)$ into the model diagram, both an isovalue $y=y_0$ and a hyper-plane $x=x_0$ are specified. The user also selects which of the parameters will be treated as the bound parameter. }

\isotrotting{} is then achieved by varying another parameter of the model, while the remaining parameters are kept fixed.
The bound parameter is automatically recomputed by a numerical solver. This ensures that the new resulting function of the model will pass through the anchor point. In our implementation, we use the binary search algorithm to find the value of the secondary parameter that satisfies the anchor point, but other search methods are possible. 

This technique proved to be invaluable for exploring the interplay of the effects of two different parameters on the model function. The inability to find a suitable value for the bound parameter under any small variation of another parameter means that the parameters are independent.



We note that this method is only suitable for exploring the multivariate parameters of the model, as opposed to complex input objects. It is, however, irrespective of the type of the underlying model computation, be it a surrogate model, or a sampled model (as defined in vPSA framework~\cite{vPSA-framework}).

\section{Design}

In this section, we describe the design of the mixed computational and visualisation solution derived from the identified requirements.

\new{Our solution allows the domain expert to analyse an empirical model and compare its results against real-world observations. 
The observed atmospheric conditions are depicted together with their theoretical corrections computed from the input parameters (\autoref{sec:air_profile}).}

\new{
The observations of the thermals are segmented and clustered from a single day of paragliding tracklogs at a given location (\autoref{sec:data}). They serve as a gallery of possible convective profiles during a given day for the domain experts to evaluate their model against. The evaluation of the model is supported by one-dimensional optimization together with \isotrotting{} and enables an efficient sensitivity analysis and assessment of the model behaviour (\autoref{sec:vertical_velocity}). }

\subsection{Rate of Sink -- AR0}

The rate of sink is modelled as a statistical variable, realizations of which contribute to the observed vertical velocity of the aircraft. The observed amount of sink depends on the pilot's steering, which we have no information about and hence model as a statistical variable. The observed vertical velocity is also affected by the vertical velocity of the air. We know that the aircraft can be in one of the three modes of flight, which we model as a Gaussian mixture. We have experimented with automatic fitting of the Gaussian mixture model. However, in the end the human interaction proves invaluable in distinguishing the different modes in a histogram representation of the vertical velocity. The estimation of the rate of sink is done once per dataset and is therefore not a burden to the user. Once the value is identified by the user, the corresponding confidence intervals are computed automatically. The user has also the advantage of comparing the value with the manufacturer's information about the aircraft.

\begin{figure}[btph]
\centering
\includegraphics[width=0.8\linewidth]{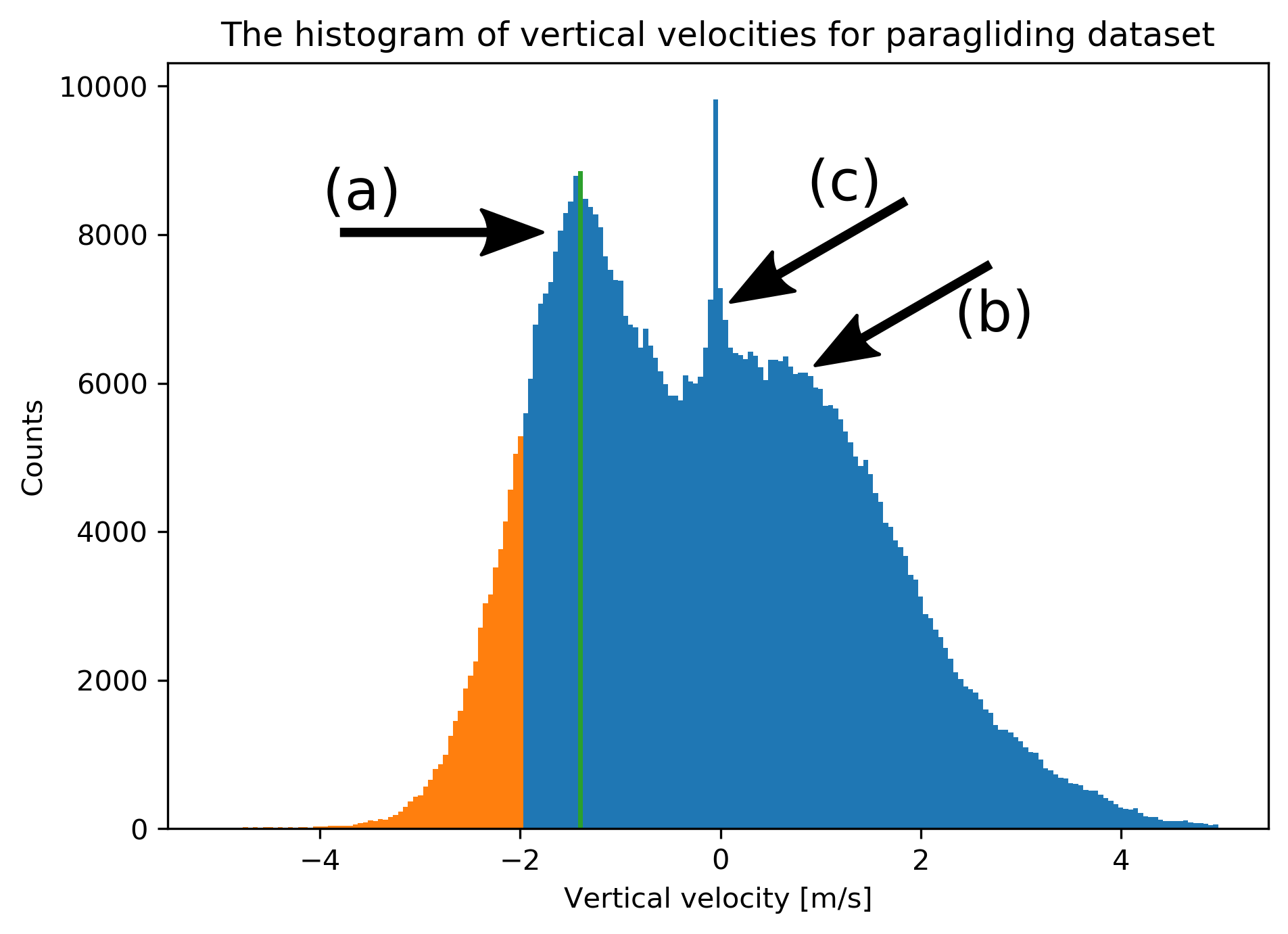}
\caption{\label{fig:sink_rate} Histogram for evaluating the sink rate. The three modes of flight are clearly visible in the histogram. The free flight on the left (a) with a one-sided 60\% confidence interval. The thermal flying in form of a Gaussian on the right (b) and a small peak around zero for soaring (c). The tall peak at zero is a no-flight mode (data acquired before takeoff and after landing). The maximum of the free flight mode is identified by the user. The estimated rate of sink for the given dataset is $(-1.37 \pm 0.55)\ \text{m/s}$.}
\end{figure}

\subsection{Paragliding Data Processing -- AR1, AR2, VR0}\label{sec:data}

In the following, we briefly comment on data processing, as well as on 3D visualization to facilitate the interactive visual analysis.  

\subsubsection{Segmentation}

The segmentation process identifies the parts of the trajectory that the pilot spent in the updraft. The characteristics of this mode of flight are positive vertical velocity and a large curvature of the trajectory. The vertical velocity is computed as the first derivative of the altitude. Theoretically, a 0 m/s of vertical velocity of the aircraft corresponds to non-zero vertical velocity of the updraft. In practice, however, the pilot will decide to leave the thermal unless experiencing a positive lift within the aircraft. 

The curvature $k$ is computed in the two-dimensional projection of the trajectory onto the surface plane as
\[
    k = \frac{x'y'' - y'x''}{(x'^2+y'^2)^\frac{3}{2}}
\]
where $'$ symbolizes the time derivative.

\subsubsection{Clustering}

Clustering identifies co-located observations. This is best done by DBSCAN~\cite{DBSCAN2017}, as it connects the observations that are within a given spatial and temporal distance, tracing the paths of trajectories being close enough to each other. The four-dimensional space-time points from the tracklogs were used as the input for the algorithm.
The parameters for DBSCAN were estimated from the average speed of the paraglider and the sampling frequency of the trajectory. 
\new{ The average speed of a paraglider is about 10 m/s; the sampling frequency of the tracking device is 1Hz; it takes about half a minute to complete a turn in the thermic flight. Based on these values, together with a short trial and error procedure the following segmentation procedure was established:}
the radius of $\varepsilon = 15 \text{m}$; and \texttt{min{\_}samples}=3, a minimum of 3 samples in a neighborhood for a point to be considered as a core point. The vertical axis was scaled by a factor of 0.2 to prevent strong convective flights to be miss-clustered.  The temporal axis was scaled by a factor of 0.5, which makes the point in close proximity of a cluster be included if it is less than 30 seconds away.

\subsubsection{3D View}

\begin{figure}[tbhp]
\includegraphics[width=\columnwidth]{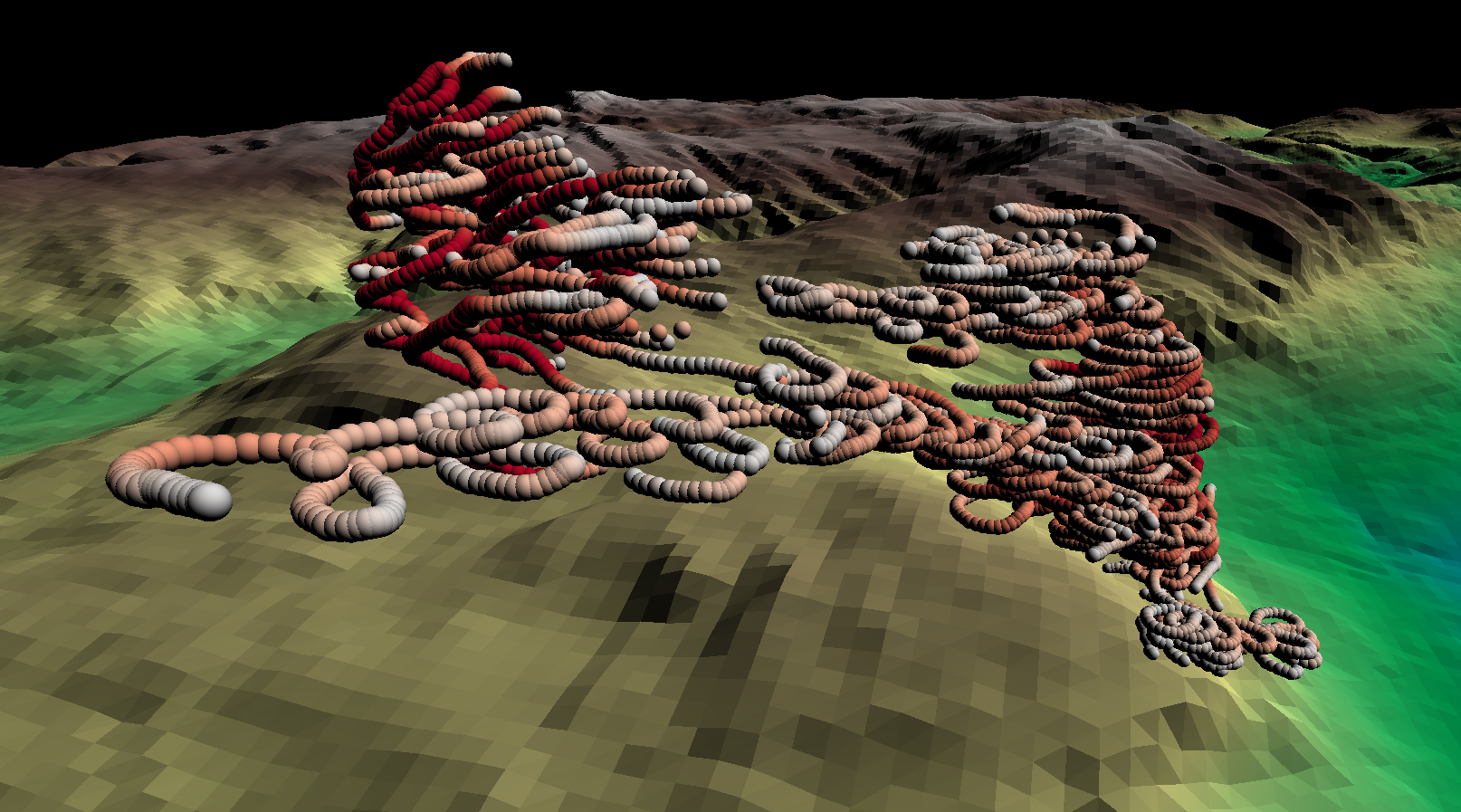}
\caption{\label{fig:3d} A 3D view of an examined thermal. The thermal has been segmented and clustered from flights of multiple pilots in the same location spanning a short time-interval of about 15 minutes. The thermal is colour-coded by the vertical velocity. The topographic map is colour-coded by altitude. In this particular case we see two clusters that were incorrectly clustered into a single thermal. The domain experts need to take this into account when modelling.   }
\end{figure}

In order to satisfy \textbf{VR0}, we provide the user with a direct rendering of the sampled trajectories that contribute to a single thermal. This enables the domain expert to consider topographical restrictions on the convective process, explore the trajectories contributing to the thermal, and evaluate the quality of the segmentation and clustering (see \autoref{fig:3d}).

The flight trajectories are colour-coded by the vertical velocity. The topographic map is colour-coded by altitude. Linking and brushing is supported for quick identification of the points between the vertical velocity profile view and the 3D view. The $x$ and $y$ coordinates of both the flight trajectories and the height data are registered using Universal Transverse Mercator projection~\cite{UTM1952}.

\subsection{Model Analysis -- AR3, AR4, VR1}

\new{To study and optimize the model, as well as the influence of its parameters, we make use of two views~-- one with a focus on the air profile information and one to facilitate IsoTrotting for parameter optimization.}

\subsubsection{Air Profile View}\label{sec:air_profile}

\begin{figure}[tb!p]
\centering
\includegraphics[width=0.7\columnwidth]{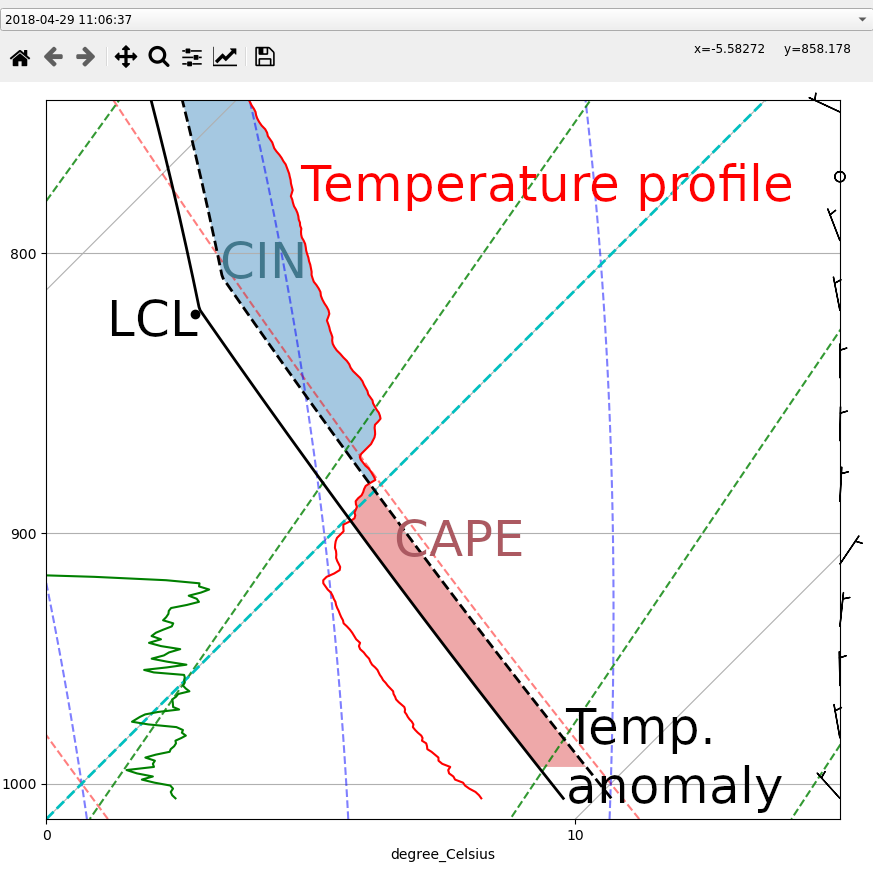}
\caption{\label{fig:metpy} The air profile view, plotted in a logarithmic scale of pressure on the y-axis skewed by 45 degrees. The x-axis is a reference for various temperature isolines: the cyan diagonal dashed line marks $0^\circ$C;
\new{the red diagonal dashed lines mark the dry-adiabatic isolines, the blue vertical dashed lines are the moist-adiabatic isolines.}
\new{The solid black line corresponds to the theoretical temperature profile} (Eq. \ref{eq:temp}), where the two-metre temperature corresponds to the position of the foot of the black line. 
The temperature anomaly is reflected in the relative position of the black dashed line with respect to the solid black line. 
The LCL is calculated from the two-metre dew point temperature and the two-metre air temperature.
The red CAPE-area, corresponds to positive buoyancy, the blue CIN-area, corresponds to negative buoyancy. 
The thermal minimum altitude is reflected in the position of the lower edge of the CAPE-area. }
\end{figure}

The model is based on the radiosonde observation of the air profile treated as a complex, environmental parameter. 
Radiosonde soundings are routinely visualized in meteorological analysis and weather forecasts~\cite{metpy}. 
We took advantage of the adopted standard and incorporated the model parameters into it, as the domain experts are already familiar with its semantics. The model itself is designed as a perturbation of the assumed state of the atmosphere, which allows some of the parameters to be directly encoded into the air sounding visualisation (see \autoref{fig:metpy}).

The air sounding is typically plotted in a logarithmic scale of pressure on the y-axis, which roughly corresponds to a linear scale in altitude. The x-axis is dedicated to temperature with the plot being skewed by 45 degrees, as the temperature rapidly decreases with altitude. The diagonal, cyan, dashed line marks 0 degrees Celsius.
\new{The temperatures can also be compared using the red, diagonal, dashed lines marking the potential temperature isolines corresponding to the dry-adiabatic process, and the blue, vertical, dashed lines that mark potential temperature isolines corresponding to the moist-adiabatic process.}

The model parameters that are based on temperature, pressure, and altitude are directly incorporated into the view. 
The surface pressure clips the observations that would fall underneath the surface level. The two-metre temperature corrects the lower part of the temperature profile and its value corresponds to the position of the foot of the black line. \new{The black line depicts the temperature profile correction based on the surface temperature computed according to Eq. \ref{eq:temp}. The black dot at the corner of the theoretical temperature profile marks the cloud base  which is computed from the surface pressure, surface humidity and surface temperature using \texttt{metpy.calc.lcl}~\cite{metpy}.}
The temperature anomaly is reflected in the relative position of the black dashed line with respect to the solid black line (surface temperature). 
The vertical velocity calculated by the model is proportional to the surface spanned between the corrected temperature profile and the perturbed temperature profile, which constitutes the buoyancy. Thus, the red area, referred to as CAPE (Convective Available Potential Energy), corresponds to positive buoyancy, whereas the blue area labelled CIN (Convective Inhibition) corresponds to negative buoyancy. 
The thermal minimum altitude is reflected in the position of the lower edge of the CAPE area.
The drag and entrainment coefficients \new{cannot be encoded in a temperature-pressure diagram and} are therefore not visualized in the air profile view. 

The view is linked to the parameter controls and the changes in parameters are immediately reflected in the air profile view.

\subsubsection{Vertical Velocity Profiles}\label{sec:vertical_velocity}

\begin{figure}[t!]
\includegraphics[width=\columnwidth]{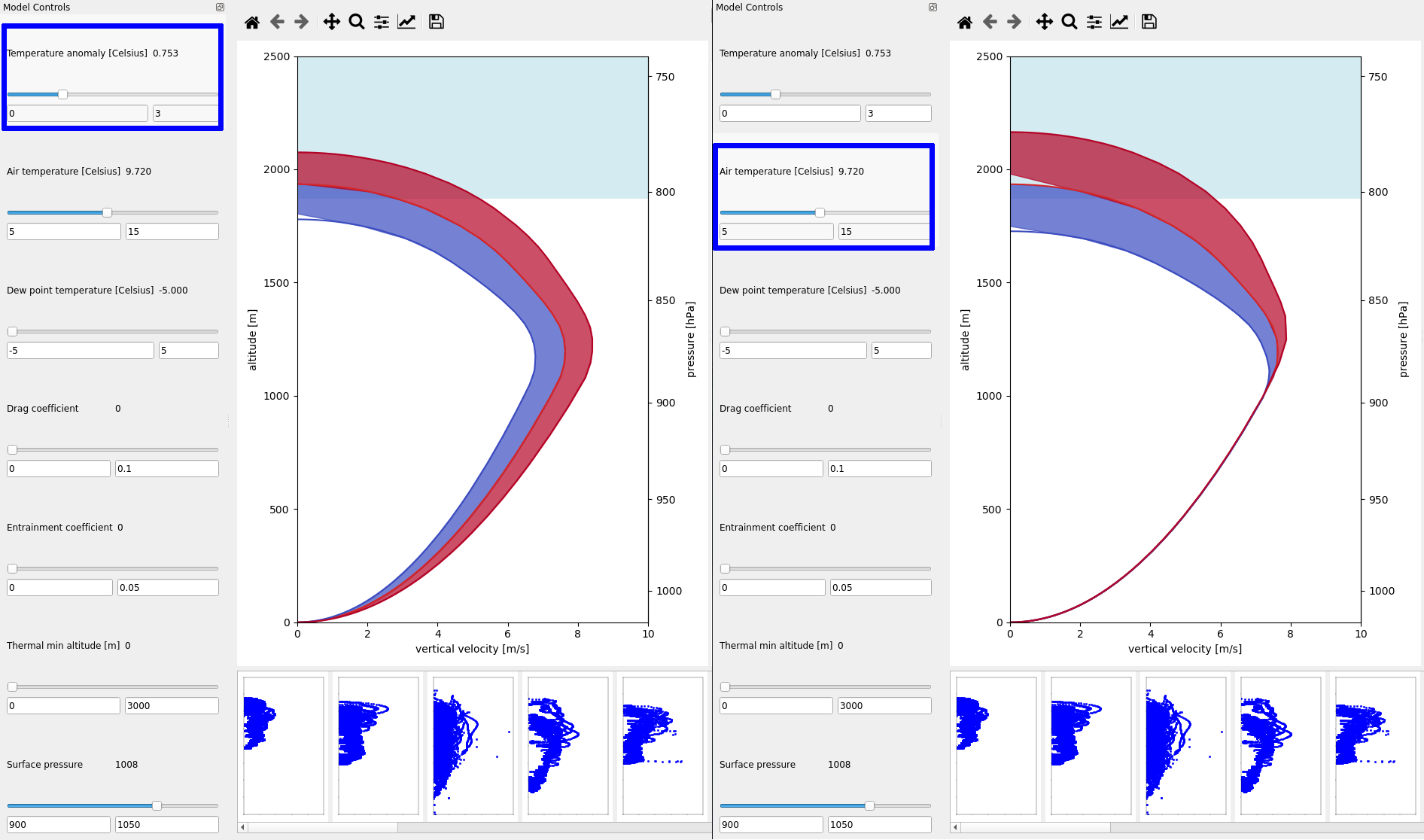}
\caption{\label{fig:hover} Variation of the model with respect to changing a selected parameter. On left the variation of temperature anomaly is shown as opposed to the variation of the surface temperature on the right. The increase of the parameter is depicted in red, the decrease in blue.}
\end{figure}

The output of a model run is a complex object called vertical velocity profile. According to \textbf{AR4}, the model run is combined with the instances of the segmented and clustered thermals in the vertical velocity profile view (see \autoref{fig:teaser}). The y-axis corresponds to the altitude and is marked with both the elevation in metres on the left hand side and the corresponding pressure on the right hand side to match the air profile view. 
The x-axis corresponds to the vertical velocity of the air at the given altitude, measured in metres per second.

The model run is depicted in orange (\autoref{fig:drag_fit}), as the theoretical vertical velocity experienced by an aircraft in the conditions described by the parameters. The plausibility of this model is evaluated against the observed data. Three versions of the observed data are overlaid with varying luminance to provide the information about the mean value and the confidence intervals.
\new{As the pilots circle in the area of the convective air, they experience the lift in a periodic manner, flying \enquote{in} and \enquote{out} of the thermal. The datapoints are shaded opaque, as the outlying shape best represents the underlying thermal, since the pilots do not experience the maximal possible lift at all time.}
Visual cues are provided for the limits of the model parameters, such as the altitude of the ground level under the  lowest measured datapoint of the thermal and the lower altitude of the clouds which corresponds to the altitude of the LCL point in the air profile view. 

As described in \autoref{sec:strategy}, we have opted for a Local-to-Global navigation strategy. 
This is achieved by visualizing a local variation of the the model with respect to a selected parameter (see \autoref{fig:hover}). A parameter is selected by hovering over the corresponding control area. The effect of a small increase of the parameter is shaded in red, whereas the decrease is shaded in blue. This kind of navigation takes care of the \textbf{AR4} -- sensitivity requirement and is aligned with the second objective, as the user obtains an immediate response for the behaviour of the model. 

As means of interaction, we provide a one-dimensional automatic regression search along a selected parameter to fit a desired value in the output space. This is our starting point to then continue with \isotrotting{}, using the selected value as the anchor point. \isotrotting{} is also a Local-to-Global navigational strategy, which helps the user to navigate to the nearest point in the parameter space, which shares the desired properties with the current run and deviates in others.

\begin{figure}[tbhp]
\includegraphics[width=\columnwidth]{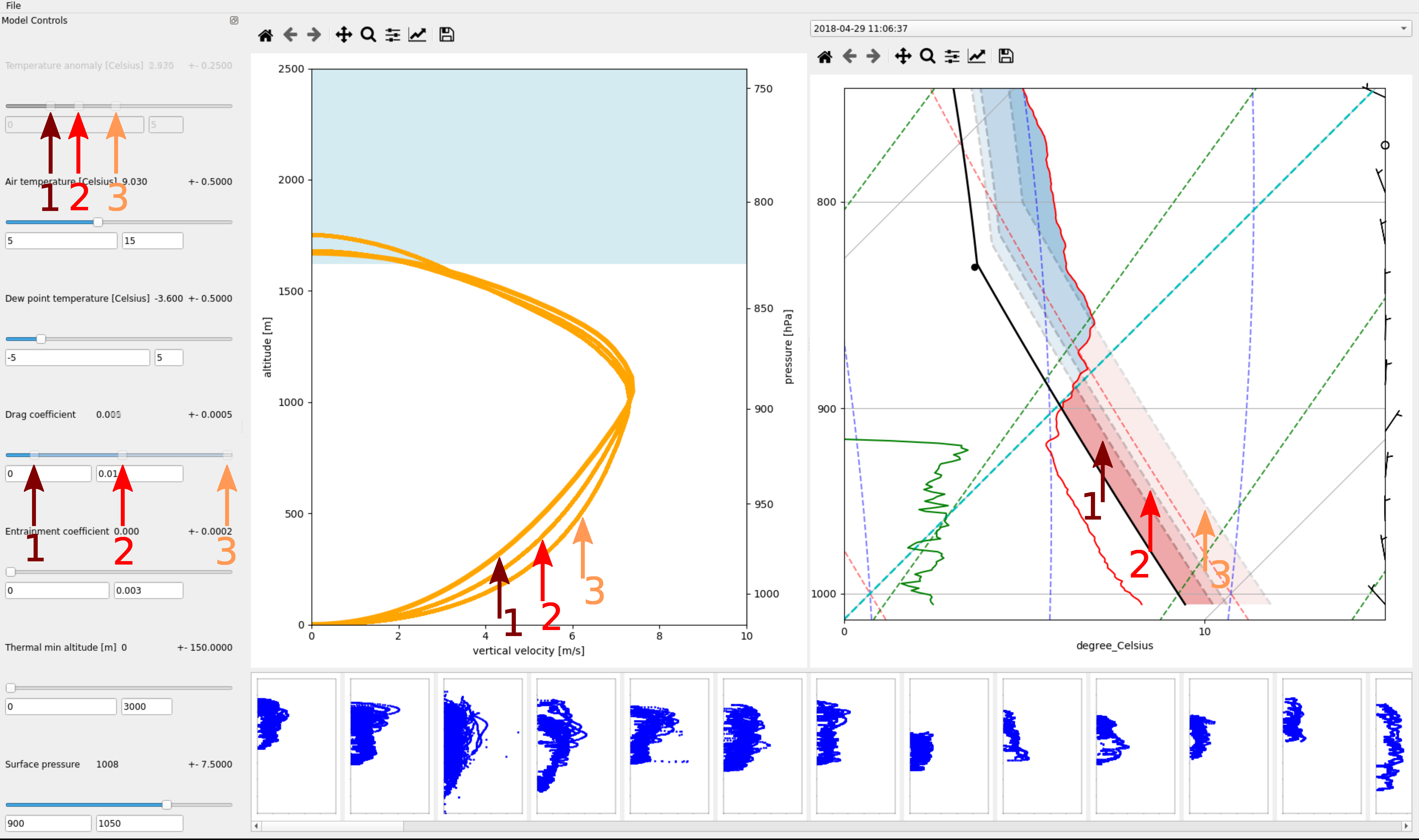}
\caption{\label{fig:drag_temp} Demonstration of \isotrotting{} in the drag and temperature anomaly parameters. The temperature anomaly is the bound parameter automatically recomputed on changing any other parameter. In this example we adjust the drag parameter value to positions marked by 1, 2 and 3, the value of the temperature anomaly is adjusted automatically, the shape of the profile changes, however the anchor point keeps the shape in place. We can observe a bulging of the shape in the lower parts of the profile.}
\end{figure}

%

%
%
\section{Application and Assessment}

In this section, we explain how our approach led to a better understanding of the empirical model and its limitations while working with a dataset.
The domain expert developed the model based on physical principals and formulated the equations provided in the supplementary material.
The model implementation was a joint effort between the visualization researchers and the domain expert. 
The use case demonstration focuses on the model validation process, even though early model visualisation helped with model implementation and debugging. 
The validation was undertaken on a paragliding flight trajectory from a competition day of the Norwegian national championship held at Voss on 29 April 2018. The dataset consists of 78 flight tracks of pilots that were competing against each other on an agreed-upon course. The area is familiar to both the domain expert and the visualization researchers.

The collaboration was carried out in repeated sessions over the course of four months. In the first phase of rapid prototyping, the visualization researchers prepared a set of visual prototypes based on agreed combination of computational analysis and visualization techniques. At the end of the prototyping phase, a visual interactive tool was built based on a few selected prototype visualizations. The tool was then jointly validated and the model performance was estimated on a provided dataset. In the following sections, we document the key insights that were obtained by the mixed machine-human analysis.

\subsection{Results}
\label{sec:results}

The domain expert had a prior intuitive and qualitative understanding of the model that was built. The key principles could be summarized as follows: 1.) The larger the temperature anomaly, the stronger the convection; 2.) Both drag and entrainment weaken the updraft; 3.) Strong updrafts are more likely on warm days.

\subsubsection{Surface Temperature Influence}
 \label{sec:res1}
The first result was derived when exploring the model itself, prior to comparing it with any observation data. Using the sensitivity analysis of the model illustrated in \autoref{fig:hover}, the domain expert observed that changing the surface temperature has seemingly no effect on the shape of the profile in the lower part (almost all the way until the maximum is reached). This qualitative observation led to a more focused examination of the profile view.

\begin{figure}[t!]
\includegraphics[width=\columnwidth]{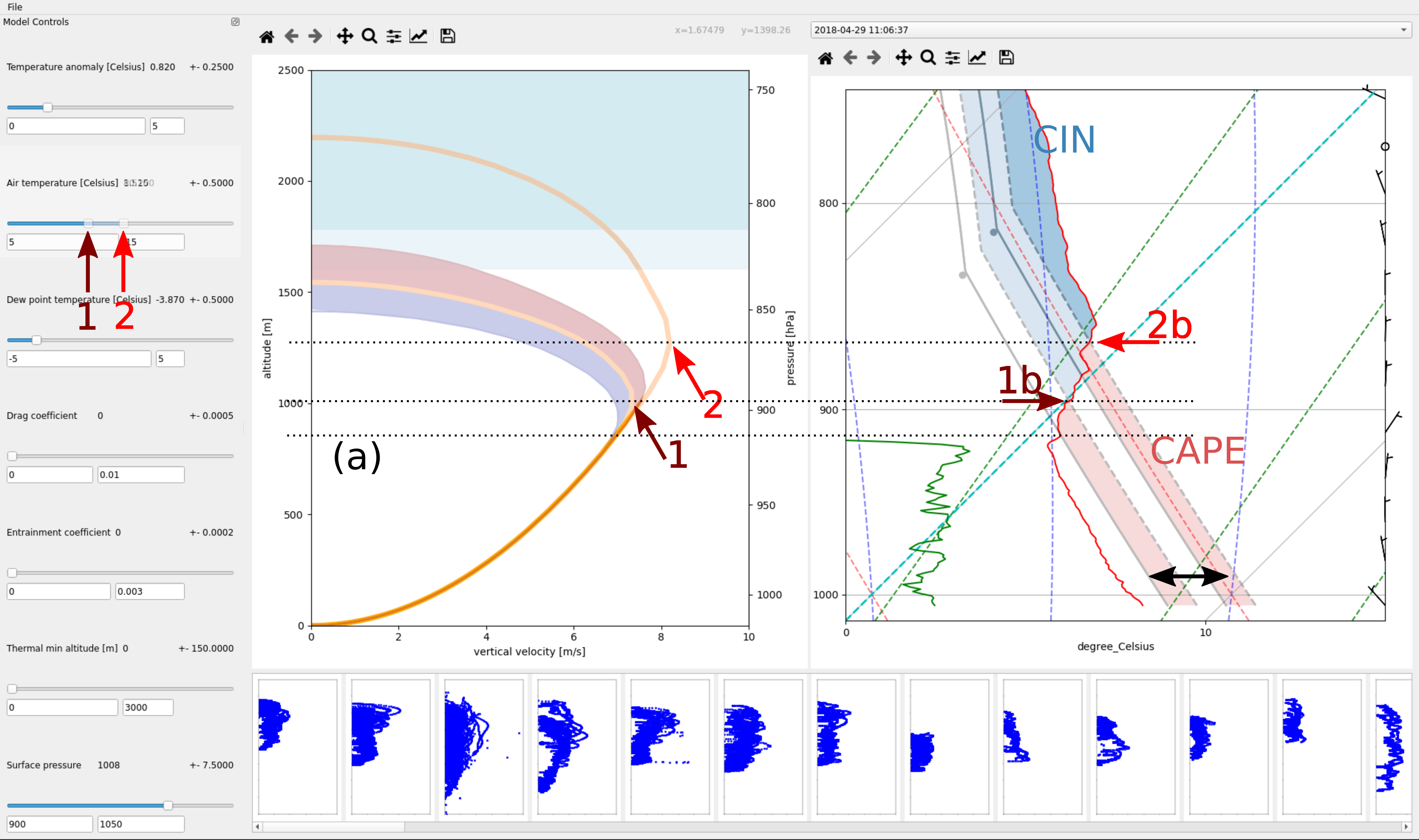}
\caption{\label{fig:air_temp} Illustrating the qualitative and principal understanding of the model behaviour. On the left, runs of the model with different air temperatures reveal the qualitative behaviour: The warmer the air, the higher the profile, with the exact same bottom part. On the right, a more principal understanding of the model: The warmer air parcel will intersect the temperature profile at a higher altitude, which results in a weaker CIN section, yielding a velocity profile that reaches higher. The identical bottom part (a) of the CAPE area results in an identical bottom part of the velocity profile.  
}
\end{figure}

Interacting with the controller for the surface temperature parameter, the user can follow a real-time update of the computed buoyancy (CAPE) profile (see \autoref{fig:air_temp}). With a few sweeps up and down the range, an invariant in the visualization becomes apparent. The shape of the profile of the CAPE region, spanned by two parallel lines with a constant gap, is moving left and right as a whole (see the video in the supplementary material). This constant gap corresponds to the constant setting of the temperature anomaly. This led to the conclusion that was confirmed by the domain expert: \enquote{\emph{Indeed, the resulting velocity profile does not depend on the the position of the CAPE region, only its area is responsible for the resulting shape.}} 

Furthermore, as the CAPE region moves left and right, the intersection point between the CIN area and the CAPE area moves along the observed temperature profile (1 and 2 on the right in \autoref{fig:air_temp}). 
Depending on the position of the intersection, the size of the CIN area will change, while the CAPE area will have the same width, but its total height will change.
One concludes that if there is a higher surface temperature with the same anomaly, the initial acceleration of the air will be the same, therefore the lower part of the velocity profile will be the same, up until the altitude where it intersects the temperature profile. With a higher temperature the intersection will occur at a higher altitude, with a reduced CIN area, yielding a velocity profile reaching to higher altitudes and achieving a higher maximum (as observed on the left in \autoref{fig:air_temp}). This fundamental understanding is even transferable between different days with different shapes.

Evaluating the model with several different air profiles measured on different days, we have confirmed that the shape of the profile of the predicted vertical lifts depends mostly on the position of the CAPE---CIN intersection. We have also confirmed that the shape of the velocity profile for a constant temperature anomaly has the same shape until the point where the CAPE area intersects the observed temperature profile. The position of the intersection, however, changes dramatically resulting in different overall shapes of the velocity profiles.

The following advice on workflow was derived from this understanding: One should first adjust the temperature anomaly based on the shape of the lower part of the profile, as one knows that the surface temperature is not going to affect it. Once the temperature anomaly has been satisfactorily set, one tweaks the surface temperature to match the position of the maximum. 
 
\subsubsection{Drag vs. Temperature Anomaly}
The next result was motivated by interactions with the observations. The database of observed convection profiles contains segmented and clustered pieces of tracklogs as explained in \autoref{sec:data}. The user selects an instance of a thermal from the bottom list and is free to explore its shape in the 3D view. The 3D view provides a sanity check whether the velocity profile can be trusted. Following the analytical steps derived earlier in \autoref{sec:res1}, the user performs a series of steps to obtain a fit of the observed thermal. First, the user selects the suspected maximum of the observed vertical velocity, where the user places an anchor point. The temperature anomaly is automatically fitted such that the model passes through the anchor point. Next, the surface temperature is adjusted such that the maxima of both the profile and the observation align, where the \isotrotting{} ensures that the profile passes through the anchor point at all times.
Thereafter, the thermal minimum altitude is adjusted such that the bottom part of the model corresponds to the observed data.

\begin{figure}[t!]
\centering
\includegraphics[width=\columnwidth]{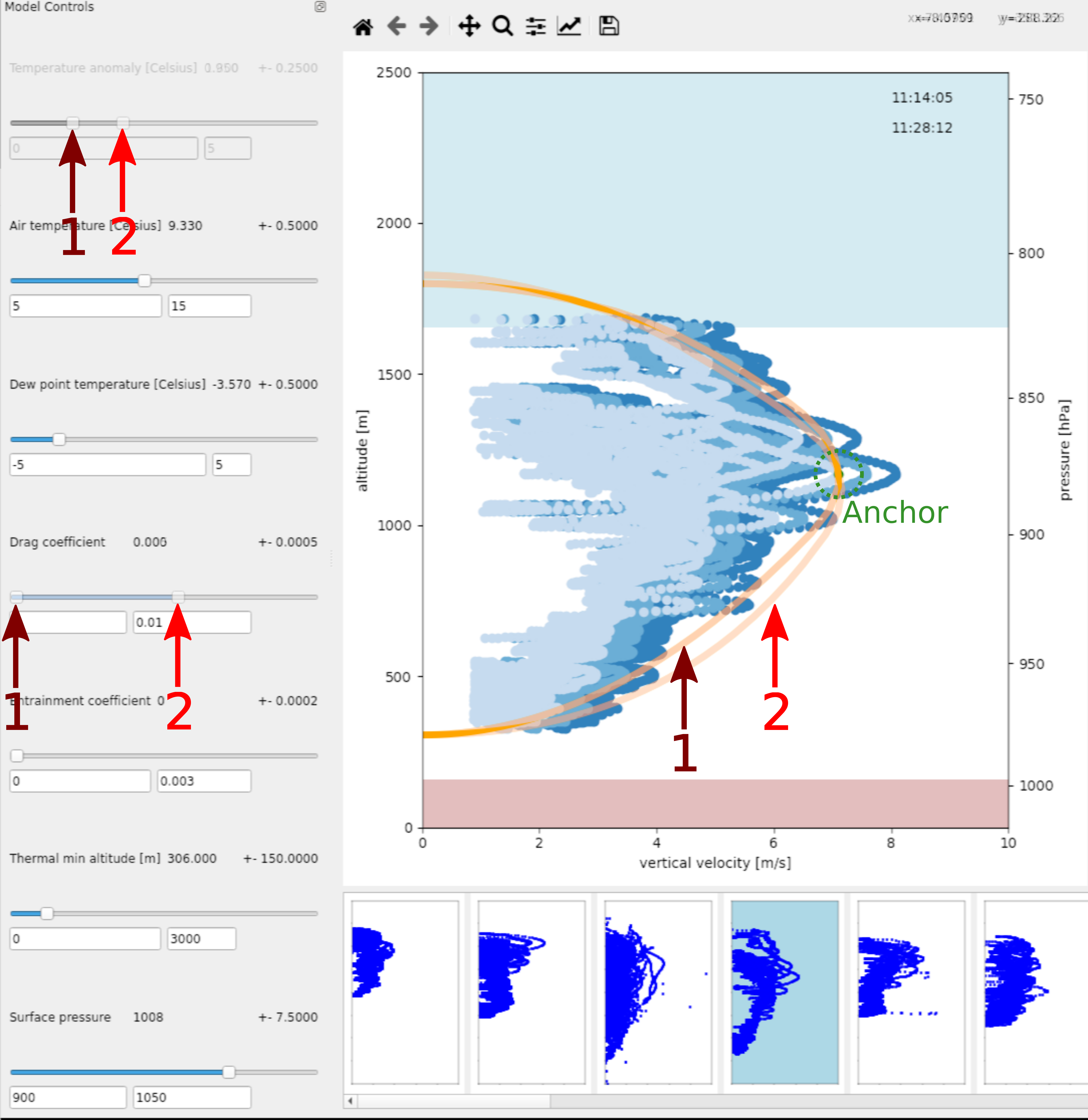}
\caption{\label{fig:drag_fit} 
The model is evaluated by placing an anchor point at the position of the suspected maximum. Adjusting the minimum altitude and the surface temperature provides a reasonable match. Changing the drag parameter does not increase the accuracy. The same is observed throughout the dataset and the estimate of the drag parameter is 0.}
\end{figure}

Using the sensitivity view, the user observes that the minimal altitude of the thermal affects both the bottom part and the top part of the model. An example of a result of such an analysis procedure can be seen on \autoref{fig:drag_fit}. As explained in the \autoref{sec:params}, all the mentioned parameters vary from one thermal instance to another. Having estimated these parameters for a selected instance, the user can explore the effects of the model parameters on the quality of the model correspondence. In this stage, the \isotrotting{} technique is invaluable, as the user can select arbitrary points along the model as a new anchor point and assign it to fix an arbitrary parameter, which allows to investigate the interplay of any two parameters on the model. A particular example would be the investigation of the effect of drag on the model run. Keeping the temperature anomaly automatically computed and adjusting the drag coefficient, the user can observe a slight change of the shape at the bottom of the profile. Performing a thorough check of the parameters, an unexpected result is discovered: the model clearly fits best with no drag. Even more unexpected, the same applies to the entrainment coefficient. 

\subsection{Domain Expert Feedback}

The proposed empirical model is the first step in the process of testing a new model for atmospheric convection. The performance of this simple model was quite good, considering that it disregards the effects of moisture in the air and has a crude description of entrainment and friction processes. An important step will be correlating the parameters of the estimated anomaly with observed values. 

A significant benefit of our approach is in the direct interaction with the parameters. Being able to immediately see the changes of the computed CAPE and CIN regions has an immense educational value.
The workflow of shaping the profile to fit the data is much better than educated guessing of the parameters. 
Being able to constraint the profile by the anchors saves a lot of parameter adjustments and makes answering the \enquote{\emph{What if}} questions much faster.

The \isotrotting{} enabled an unparalleled opportunity to investigate the interplay of two parameters on the model behaviour. Using the traditional approach of manually plotting model runs, one would never be able to reliably confirm the complex interactions with such a high efficiency.
The alternative would be to manually select two parameters, evaluate the model at discrete samples of a fixed interval and filter the outputs based on a simple criterion, such as the maximum value. The \isotrotting{} makes this much faster, enables the use of arbitrary pairs of parameters and the filtering criteria is defined by clicking on the diagram. 

\subsection{Implementation}

The rapid prototyping was done in Python Jupyter notebooks using Matplotlib~\cite{matplotlib} and Pandas~\cite{pandas}. The final application is done in PyQt5 and Matplotlib~\cite{matplotlib}. The temperature profile view is based on the \texttt{metpy}~\cite{metpy} package.

The solution runs on a system with the following parameters:~ CPU: Intel(R) Core(TM) i7-7700K @ 4.20GHz;~ RAM: 2 x 16GiB DIMM DDR4 2400 MHz;~ GPU: GeForce GTX 1080, 8 GB GDDR5X;~ Storage: 500GB SSD Samsung 860 Evo.
The initial parsing of the tracklog files and the subsequent computational processing takes circa 5 minutes. The visual analysis is fully interactive.

\section{Discussion}

This work managed to answer some important question regarding the modelled vertical velocity profiles of atmospheric convection (see \autoref{sec:results}). 
More than that, it established a powerful method for analysing pairwise parameter dependencies in visual parameter space analysis. This enables domain researchers to understand their models much faster than before.

This work is a part of an effort of building an empirical model for atmospheric convection.
Even though the analysis tool helped the domain experts to investigate low altitude atmospheric convection for a single day of observations, there are plans to do a subsequent statistical analysis on a larger dataset to obtain more comprehensive results.
The subsequent analysis will be built on the findings from this work. For example, a meaningful error metric for automatic comparison between the observed and simulated velocity profiles will need to be devised. Even then, a fully automatic solution will probably not be feasible and a combination of an automatic analysis and visual exploration will be necessary. 

\new{Several assumptions were made when designing the differential equations of the model. Fist of all, a one-dimensional model is used that assumes that all the properties inside a thermal are constant and different from the properties outside of the thermal. This assumption is necessary for a concise model. The discontinuity is, however, not real and needs to be taken into account, for example, as the paragliders enter and leave the thermal. 
Having the model one-dimensional also means that it does not take into account the transient movement of the thermal. As we are only dealing with forces that act in the vertical direction (gravitation, buoyancy), we neglect the role of transient movement in the vertical direction.
The transient movement can have consequences, which are not captured by the empirical model. However, the anticipation and the treatment of these is left to the user and supported by the 3D visualization.}

\new{
The equations also describe a steady state, i.e., not a time-dependent behaviour. This is justified by comparing the model with a single occurrence of the thermal (circa 10 minutes), during which the values are assumed unchanged. The investigation of the daily temporal behaviour is planned as future work. }

\new{The entrainment is assumed to have a linear dependency on the vertical velocity, i.e., the faster the thermal, the more mixing with the environment occurs. This is a reasonable assumption, but perhaps an oversimplification and reason why the entrainment analysis yields values of zero. }

Further limitations of the studied empirical model are due to the disregard of the effects of moisture making the results unreliable near the condensation level and above it, in the clouds. There are plans to improve the model and take the moisture into account in the future. \new{This would allow repeating the analysis with focus on the influence of moisture on the shape of the profiles, also above the cloud base.}

Limitations of the visualization tool include the possibility to work with only one dataset at a time. Loading different data is possible, but no visual support for a comparison across datasets is implemented. This corresponds to the fact that only a single dataset was available at the time of writing this paper. Similarly, no visual support is devised for comparing models of multiple days. This is not a drawback at the moment, but rather part of future work to support forecasting.

We consider the \isotrotting{} technique as a versatile tool to efficiently explore parameter spaces by making simple restrictions in the output domain and it could possibly be used in other applications, even though it focuses on scalar models only. There is a possibility to extend this technique to higher-dimensional mathematical objects, but that is beyond the scope of this work.

\section{Conclusion and Future Work}

In this paper, we presented a design study in empirical modelling of atmospheric convection in meteorology. We developed a visual parameter space analysis approach to investigate the behaviour of a new model, introducing a novel navigation strategy to address the interplay of pairs of parameters. The novel strategy allows domain experts to test complex behaviour of the model in a much more efficient manner. The validation of the empirical model is performed on a novel datasource compiled from freely available tracks of paragliding flights (\url{https://www.flightlog.org/}). The relevant measurements are extracted from the data by means of a combined computational and interactive solution. Both the uncertainties of the measurements and the sensitivity of the model are considered during the analysis. \textcolor{black}{The relevant findings of the modelling are documented and reported.}

\new{At present, the model can estimate the strength of the atmospheric convection based on readily available observations. The evaluation of the model can be done in 0.1 second, which makes it suitable as a subroutine in more complex simulations. 
A limited level of complexity of the model, together with a visual interactive application make it appropriate also for education purposes. }
\new{In the future, the model could also be used for forecasting the strength of the atmospheric convection based on the forecasts of air temperature profiles.
In order to achieve this, the model will be subject to more analysis using more observations as well as in situ measurements of meteorological parameters in the thermals.
Based on our findings, we plan to extend the empirical model to improve its performance, introduce effects of moisture as well as calculations within clouds.}

\new{Most importantly, this paper presents a novel interaction technique for visual parameter space analysis. The technique enables constraining the parameter space of a model by interacting with the model's output. This is leveraged to efficiently explore various parameter configurations supporting a desired outcome. 
All this is enabled by a rigorous mathematical treatment of the parameter space, and as such, it brings together computational analysis with interactive visualization.
In our case, this combination of approaches enables a much faster iteration through multiple hypotheses, when engaged with scientific modelling.}

\acknowledgments{
We would like to thank the Norwegian Airsport Federation for providing the paragliding data and the Norwegian Meteorological Institute for providing the meteorological observations. 
Parts of this work have been done in the context of CEDAS, i.e., UiB's Center for Data Science.  
}

\bibliographystyle{abbrv-doi-hyperref}
\bibliography{manuscript}

\include{convection-model}

\end{document}

%% file: convection-model.tex



\section*{Appendix A -- Convection Model}

We start from the equations the equations of motion for one-dimensional motion ($u=v=0$) in the vertical
\begin{eqnarray}
\frac{dw}{dt} &=& \frac{\partial w}{\partial t} +w\frac{\partial w}{\partial z} = -\frac{1}{\rho}\frac{\partial p}{\partial z} - g - \alpha w \\
\frac{db}{dt} &=& \frac{\partial b}{\partial t} +w\frac{\partial b}{\partial z} = -\beta b~,
\end{eqnarray}
where $w$ is the vertical velocity and $b=g\frac{\theta'}{\overline{\theta}}$ is the buoyancy, and $\alpha$ and $\beta$ are the respective damping parameters corresponding to linear entrainment.

Assuming a basic state such that
\begin{eqnarray}
\theta &=& \overline{\theta}(z)+\theta' \\
p &=& \overline{p}(z)+p' \\
\rho &=& \overline{\rho}+\rho' \\
w &=& w'
\end{eqnarray}
where $\overline{p}$ is hydrostatically balanced $\frac{d \overline{p}}{d z}=-\overline{\rho}g$ and there is no basic state vertical velocity $\overline{w}=0$, yields
\begin{eqnarray}
\frac{\partial w'}{\partial t} +w'\frac{\partial w'}{\partial z} &=& -\frac{1}{\overline{\rho}}\frac{\partial p'}{\partial z}-g\frac{\rho'}{\overline{\rho}} - \alpha w' \\
\frac{\partial b'}{\partial t} + w'\frac{\partial b'}{\partial z} + w'\frac{d \overline{b}}{d z} &=& -\beta b'~,
\end{eqnarray}
where we assumed $p'\ll\overline{p}$ and $\rho'\ll\overline{\rho}$.

Assuming steady state $\frac{\partial}{\partial t}=0$ and that the perturbation pressure produced by the thermal is negligible $p'=0$, which is common with convective parcel theory, we obtain
\begin{eqnarray}
w'\frac{\partial w'}{\partial z} &=& b' - \alpha w'  \label{eq-w} \\
w'\frac{\partial b'}{\partial z}  &=& - w'\frac{\partial \overline{b}}{d z}-\beta b'  \label{eq-b}~,
\end{eqnarray}
where we assumed the Boussinesq approximation $-g\frac{\rho'}{\overline{\rho}}=g\frac{\theta'}{\overline{\theta}}=b'$. If we assume that the buoyancy entrainment has a proportionality to the vertical velocity, $\beta=\gamma w'$, we obtain
\begin{eqnarray}
w'\frac{\partial w'}{\partial z} &=& b' - \alpha w \label{eq-w2} \\
\frac{\partial b'}{\partial z}  &=& - \frac{d\overline{b}}{d z}-\gamma b'  \label{eq-b2}~.
\end{eqnarray}
For the convectively mixed layer ($\frac{d \overline{b}}{d z}=0$) we can solve for $b'$
\begin{eqnarray}
\frac{\partial b'}{\partial z}  &=& -\gamma b'~,
\end{eqnarray}
yielding
\begin{eqnarray}
b'  &=& b'_{z=0}~e^{-\gamma z}~.
\end{eqnarray}
This solution can be used to solve
\begin{eqnarray}
w'\frac{\partial w'}{\partial z} &=& b'_{z=0}~e^{-\gamma z} - \alpha w'~,
\end{eqnarray}
which you can probably look up in a mathematics solution book or try to solve analytically yourself. In the simple case of $\gamma=\alpha=0$, the problem reduces to
\begin{eqnarray}
w'\frac{\partial w'}{\partial z} &=& b'_{z=0}~,
\end{eqnarray}
which can be rewritten as 
\begin{eqnarray}
\frac{\partial}{\partial z}\left(\frac{1}{2}w'^{2}\right) &=& b'_{z=0}~,
\end{eqnarray}
and yields $w'=\sqrt{2~b'_{z=0}~z}$. In fact, using (\ref{eq-w}) with $\alpha=0$ yields
\begin{eqnarray}
\frac{\partial}{\partial z}\left(\frac{1}{2}w'^{2}\right) &=& b',
\end{eqnarray}
which is used to calculate the convective available potential energy ($CAPE$)
\begin{eqnarray}
CAPE(z) &=& \int_{z=0}^{z}b'~.
\end{eqnarray}

The interesting part of the problem is when one also allows for $\frac{d \overline{b}}{d z}>0$ above the mixed layer, i.e., you allow for a capping inversion that stops the convection. In that case you would need to solve (\ref{eq-w}) together with (\ref{eq-b}) or (\ref{eq-w2}) together with (\ref{eq-b2}) numerically anyway.
